\documentclass[preprint,12pt]{elsarticle}

\usepackage{amssymb}
\usepackage{url}
\usepackage{array}
\usepackage{xcolor}
\usepackage{siunitx,subfigure}

\graphicspath{{figs/}}

\journal{Nature}

\begin{document}

\begin{frontmatter}

\title{Pavlovian reflex in colloids}

\author[uwe]{Noushin Raeisi Kheirabadi}
\ead{Noushin.Raeisikheirabadi@uwe.ac.uk}
\author[iit,uwe]{Alessandro Chiolerio}
\author[uwe]{Andrew Adamatzky}


\affiliation[uwe]{organization={Unconventional Computing Laboratory, UWE, Bristol, BS16 1QY, UK}}

\affiliation[iit]{organization={Center for Bioinspired Soft Robotics, Istituto Italiano di Tecnologia,  Genova, Italy}}

\begin{abstract}
Pavlovian reflex is an essential mechanism of nervous systems of living beings which allows them to learn. Liquid colloid computing devices offer a high degree of fault-tolerance, reconfigurability and plasticity. As a first step towards designing and prototyping colloidal neuromorphic computing systems we decided to evaluate if it is possible to implement Pavlovian reflexes. We equate an increase of a synaptic weight with decreased resistance.
In laboratory experiments we demonstrated that it is possible to implement Pavlovian learning using just two volumes of colloid liquid. 
\end{abstract}




\end{frontmatter}

%
%
%

\section{Introduction}
\label{sec:sample1}

Liquid computers have been known since 1990s when Arnold Emch demon\-strated hydraulic extraction of  the $n$\textsuperscript{th} root of any number~\cite{emch1901two}. Later prototypes of liquid computers included hydraulic integrators~\cite{moore1936hydrocal,luk1939hydraulic}, monetary national income analog computers~\cite{bissell2007historical}, fluid mappers~\cite{moore1949fields}, fluid logics~\cite{hobbs1963fluid}, Belousov-Zhabotinsky computers~  \cite{kuhnert1989image,agladze1996chemical,steinbock1996chemical, gorecki2003chemical,adamatzky2005reaction}, reaction-diffusion computers\cite{adamatzky1996reaction,tolmachiev1996chemical,adamatzky1997chemical,adamatzky1994constructing}, fluid maze solvers~\cite{fuerstman2003solving}, 
flow driven droplet logics~\cite{cheow2007digital,fair2007digital,toepke2007microfluidic,prakash2007microfluidic}, liquid marbles logic~\cite{draper2017liquid}, see detailed overview in \cite{adamatzky2019brief}. In these devices a liquid is used to represent signals, actuate mechanical computing devices and to modify signals via chemical reactions. Most prototypes of liquid computers are specialised processors, capable for solving tasks in very narrow domains. The most natural application of a liquid computer is for controlling liquid autonomous systems \cite{chiolerio2017smart}. To generalise applications domains of liquid computers we should embed them into neuromorphic architectures, including learning abilities~\cite{kheirabadi2022learning}. In \cite{raeisineuromorphic} we provided an exhaustive analysis of experimental laboratory prototypes where neuromorphic systems are implemented in liquids, colloids and gels. The analysis shown that whilst there are implementations of artificial synapses in liquid phase no cascaded liquid neuromorphic systems have been prototyped. To fill the gap we conducted a series of laboratory experiments on Pavlov's conditional reflexes~\cite{pavlov1906scientific,pavlov1928lectures} in colloids and produced a first ever prototype of Pavlovian reflex liquid circuit. 

The paper is structured as follows. Section~\ref{methods} introduces an experimental setup. Functioning of the colloid Pavlovian circuit is analysed in Sect.~\ref{results}. Pathways to integration of Pavlovian colloid circuits into liquid neuromorphic systems are discussed in Sect.~\ref{discussion}.

\section{Methods}
\label{methods}

Surfactant solution with a concentration of 0.22 wt\% was prepared by adding sodium dodecyl sulphate (SDS, Merc) in de-ionised water (DIW, prepared in the lab with Millipore de-ionised water generator device, model Essential, rated 15 MOhm cm) and stirred to get a homogeneous solution. 1 mg ZnO nanoparticles  (US Research Nanomaterials, Inc.) were added to DMSO  (US Research Nanomaterials, Inc.)  with continuous stirring. Then 2 ml of SDS solution and 1 ml NaOH 10M added to the mixture under stirring. The concentration of resulting dispersion was maintained at 0.11 mg/ml. The resulted suspension was placed in an ultrasonic bath for 30 minutes. Then stirring process was continued for a few more hours to get a uniform dispersion of ZnO~\cite{anand2017role}.

All electrical measurements were made with Fluke 88464A and Keithley 2400. Field emission scanning Electron Microscopy (FEI Quanta 650 FESEM) was also used to characterise the nanoparticle suspensions. The accelerating voltage was set to 10 kV in this study, and the working distance was set to around 5 mm. The images' contrast and brightness were optimised so that particles could be easily distinguished from the background.
Ultraviolet-visible (UV-vis.) spectrometer (Perkin Elmer Lambda XLS) was used to measure sample absorbance at room temperature.

\begin{figure}[!tbp]
    \centering
    \subfigure[]{\includegraphics[width=0.49\textwidth]{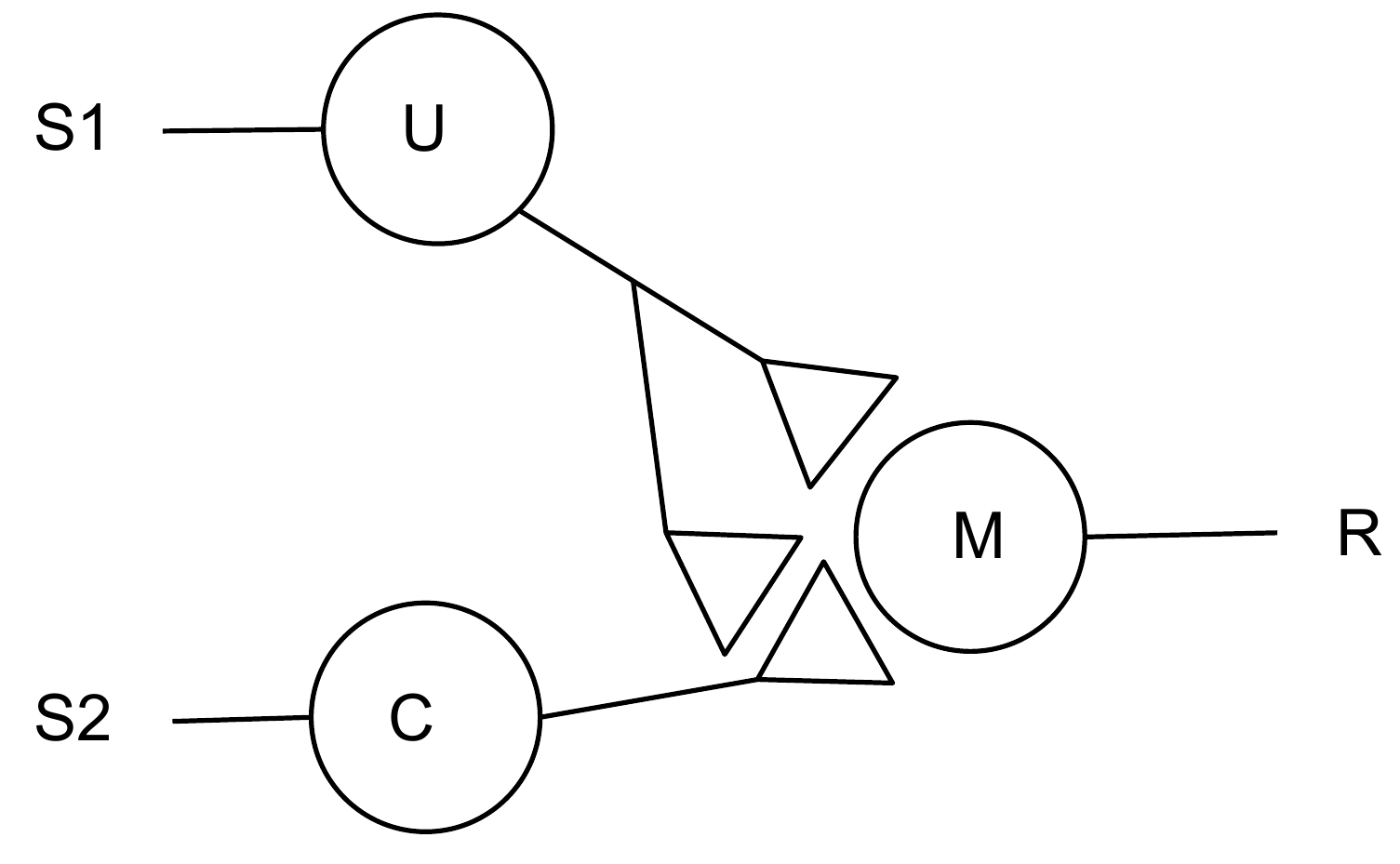}} 
    \subfigure[]{\includegraphics[width=0.49\textwidth]{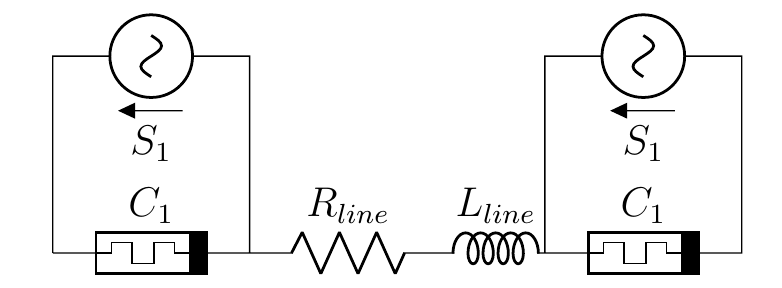}} 
    \subfigure[]{\includegraphics[width=0.49\textwidth]{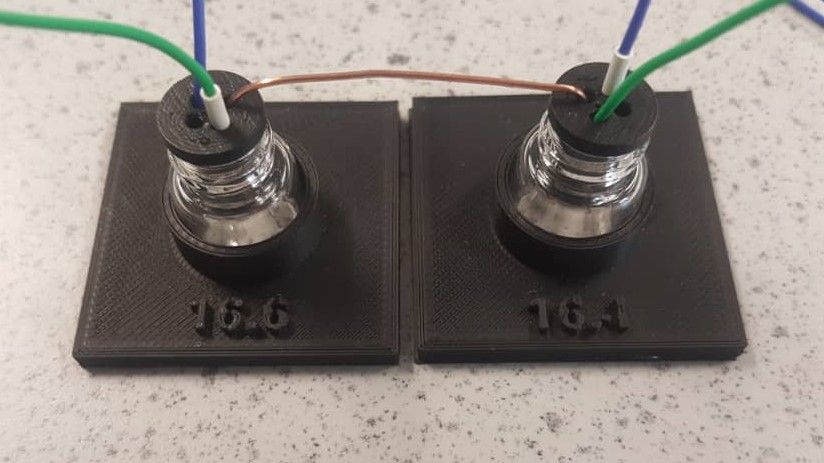}}
    \caption{A scheme of Pavlovian learning: (a)~abstract neurons scheme, S1 is unconditional stimulus, S2 is conditional stimulus, U is unconditional neuron, C is conditional neuron, M is a motor neuron, synapses are shown as triangles, 
    (b)~electronic real circuit, where the connection wire between samples is modeled as a unique transmission line by a resistor and an inductor in series, the two colloidal suspensions by two memristors and the input signals by a voltage source; (c)~Photo of the experimental setup.}
    \label{fig:scheme}
\end{figure}

To examine the colloid's ability to learn the Pavlovian learning algorithm was chosen, which has four steps: ringing bell, feeding food, reaching salivation, repeat the cycle. 
Based on Pavlov's famous experiment~\cite{pavlov1906scientific,pavlov1928lectures}, ringing the bell by itself will not cause salivation. On the other hand, salivation appears for food. The dog is repeatedly fed after ringing the bell during the training procedure, which establishes an association between food and ringing. A ringing bell can cause salivation after such training, as shown in condition. The potential mechanism of such learning is shown in Fig.~\ref{fig:scheme}. The axon from a neuron U reacting to unconditional stimuli branches out to establish synaptic connections both on the motor neuron M and synaptic connection from the conditional neuron C to M (Fig.~\ref{fig:scheme}a). When neuron U fires the action potential propagates to M and excites but also propagates through synapse C$\rightarrow$M thus increasing its synaptic weight (decreasing the synaptic resistance). In designing the colloid Pavlovian circuits our assumptions were that at the onset of learning resistance of a colloid synapse C$\rightarrow$M is so high that current passing to M is not high enough to excite M. Simultaneous stimulation of colloids U and C leads to decrease of resistance in C$\rightarrow$M to such a degree, that when C is stimulated alone the current reaching M is enough to excited M. This is expressed in an electronic scheme as shown in Fig.~\ref{fig:scheme}b. The electronic scheme was converted to experimental setup as shown in Fig.~\ref{fig:scheme}c.

\section{Results}
\label{results}

\subsection{Colloid Structural Characteristics}

\begin{figure}[!tbp]
\centering
\subfigure[]{
\includegraphics[width=0.7\textwidth]{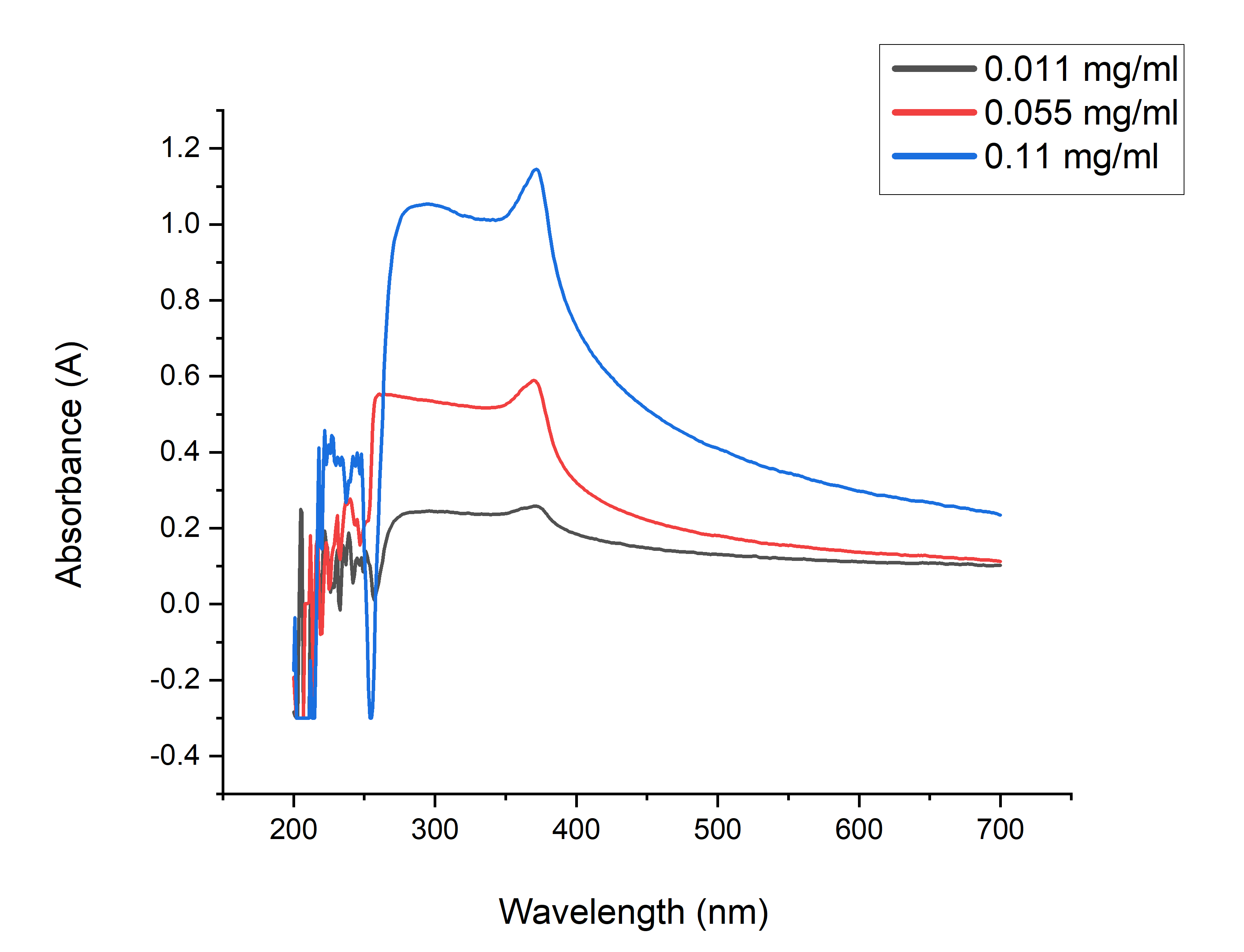}\label{fig:UV}}
\subfigure[]{\includegraphics[width=\textwidth]{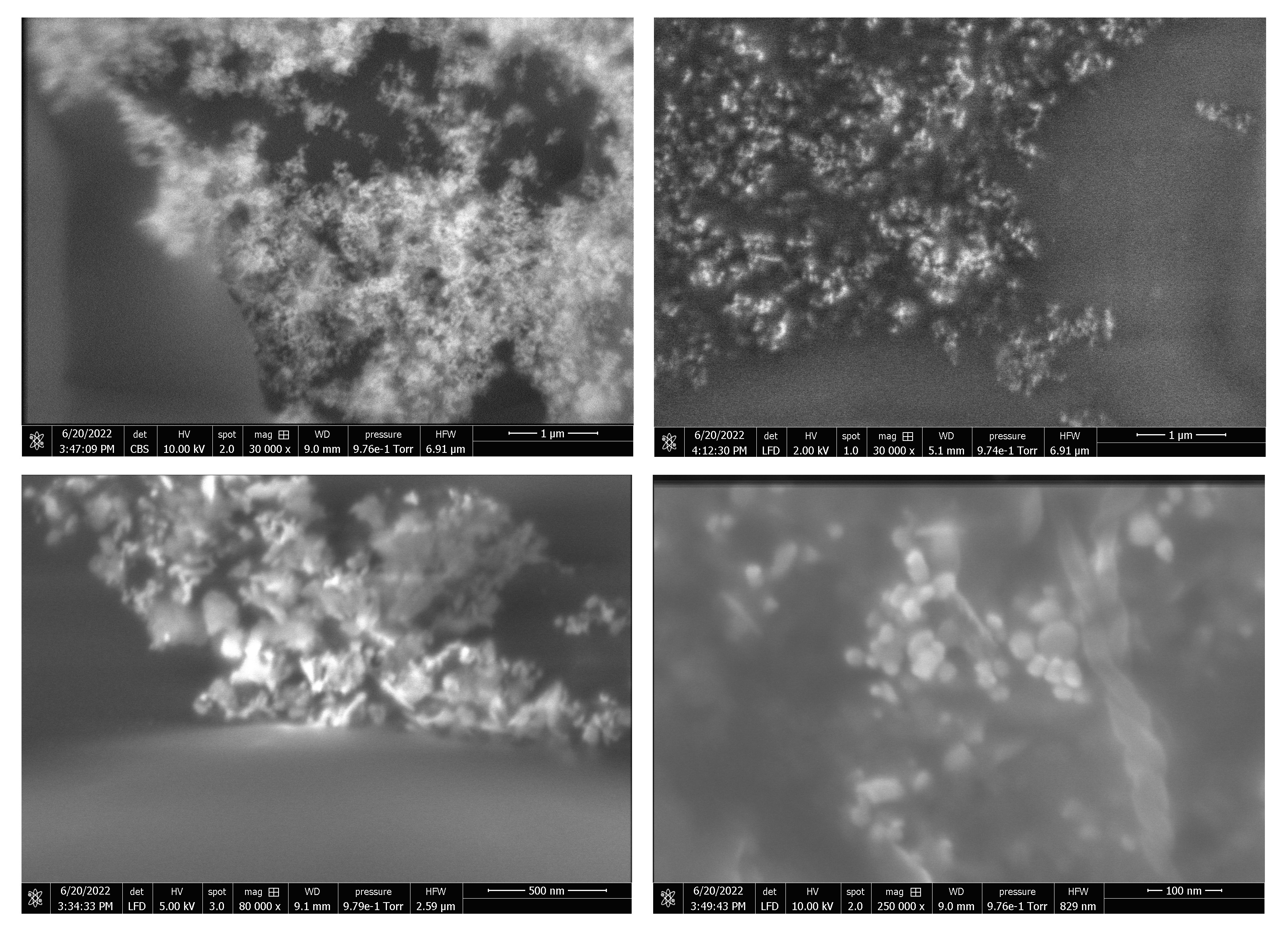}\label{fig:SEM}}
\caption{Structural characterisation of colloid used in experiments. (a)~UV-Visible spectra of the sample with concentrations 0.011, 0.055, and 0.11 mg/ml. (b)~SEM images of drop casted ZnO colloids on Copper substrate in different magnifications}
\label{fig:characterisation}
\end{figure}


At room temperature, UV–visible absorption spectra of ZnO colloids at three different concentrations are measured in the wavelength range of 200–700 nm. The plots of UV-Visible absorption spectrum are shown in Fig.~\ref{fig:UV}. Three peaks in the spectrum are observed at 370~nm to 371~nm for the colloid concentration of 0.011 mg/ml, 0.055 mg/ml, and 0.11 mg/ml. The peaks at 370~nm to 371~nm are the characteristic absorption peaks of hexagonal ZnO nanoparticles~\cite{pudukudy2015facile}. According to previous reports~\cite{reddy2011combustion,sun2011enhanced}, the results are in good agreement with bulk ZnO (370 nm). Calculation of the optical band gap was based on the below equation:

$$
E_g (eV) = hc/\lambda = 1240/\lambda
$$

Where $E_g$ is the optical band gap, h is the Planck’s constant, c is the speed of light,and $\lambda$ is the wavelength of maximum absorption. Here, the optical band gap was calculated as 3.35 eV, which is consistent with earlier studies~\cite{reddy2011combustion,baskoutas2010conventional,baskoutas2011transition}.

The morphology and size of particles were studied by FESEM using a thin layer of ZnO colloid prepared by drop casting a drop of as-synthesized colloid (0.11 mg/ml) onto a 100 $\mu$m thick copper foil at room temperature. Fig.~\ref{fig:SEM} illustrates FESEM results showing agglomeration of particles during sample preparation. FESEM images rarely reveal separated ZnO spheres due to the surface tension of the solvent during evaporation, and most of the ZnO spheres are multilayered. As a result, the nanoparticles would re-aggregate during drying since increased liquid surface tension would draw them closer together~\cite{lu2018methodology}.

\subsection{Colloid Learning}

\begin{figure}[!tbp]
\subfigure[]{\includegraphics[width=0.49\textwidth]{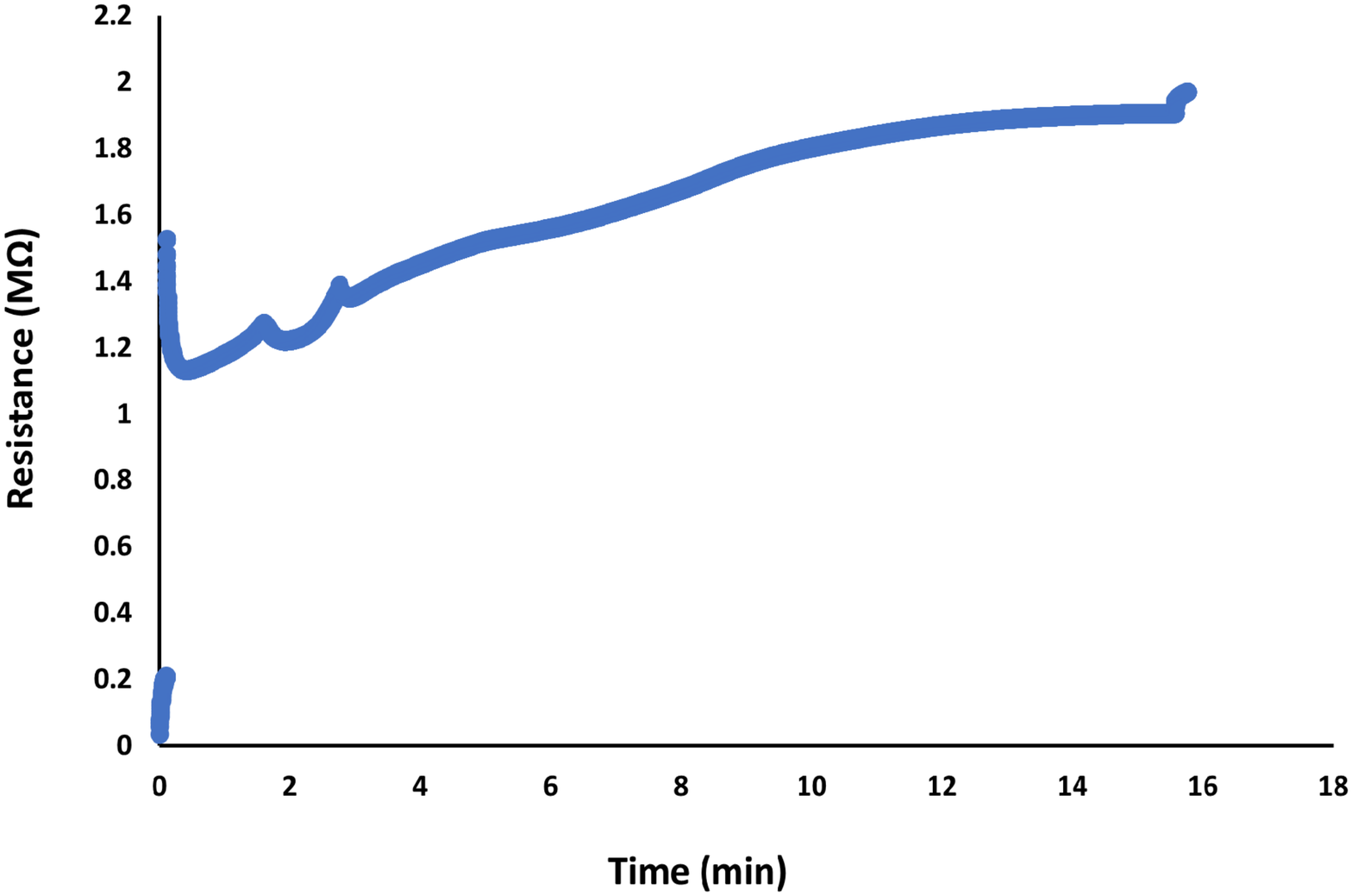}}
\subfigure[]{\includegraphics[width=0.49\textwidth]{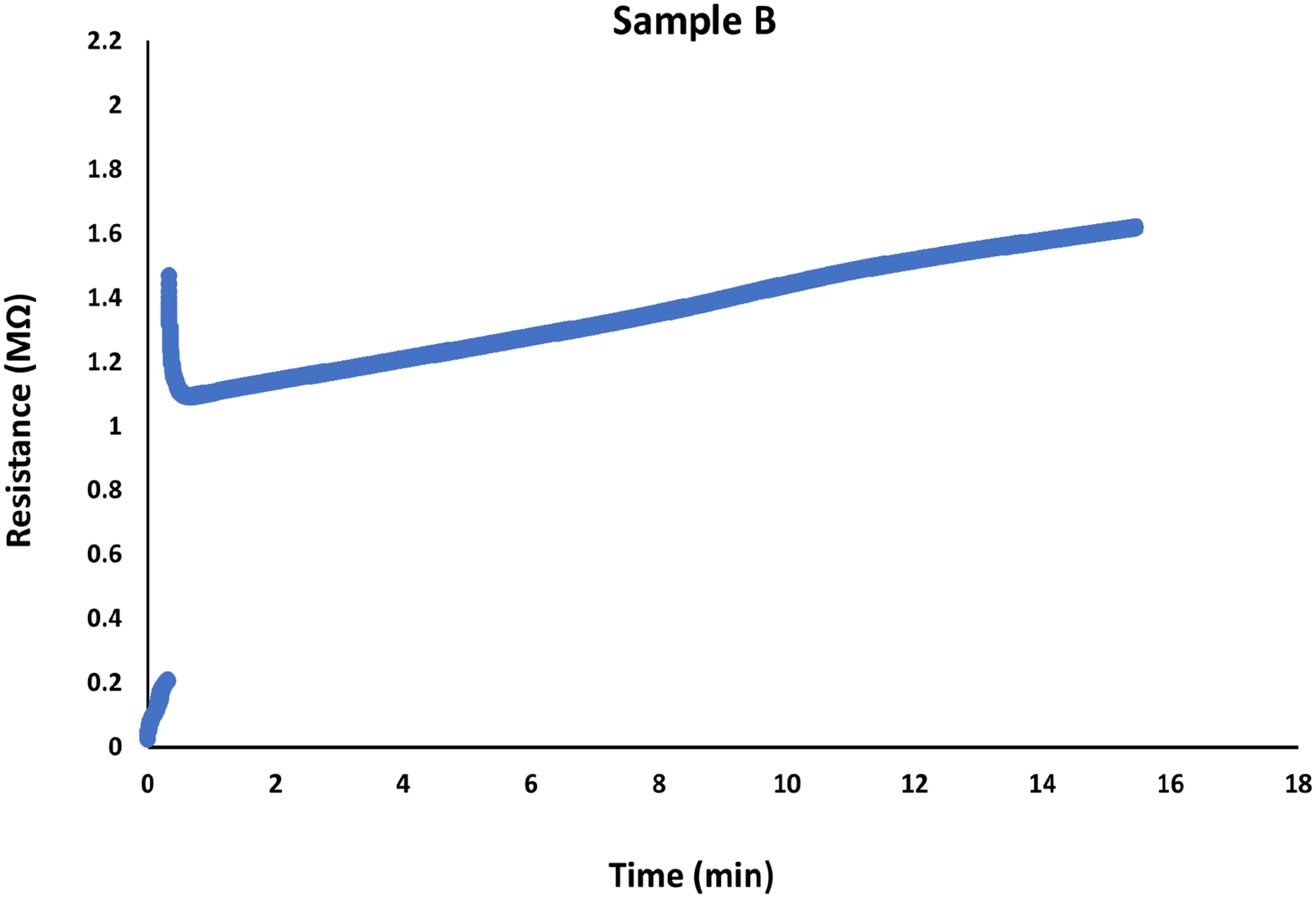}}
\caption {Resistance diagram of sample A and B, before applying stimulation}
\label{fig:before}
\end{figure}

The Pavlov’s dog experiment was mimicked using a synthesised colloid. To perform Pavlovian learning, first, the resistance of sample (A) and sample (B) were measured (Fig.~\ref{fig:before}). 

\begin{figure}[!tbp]
\subfigure[]{\includegraphics[width=0.49\textwidth]{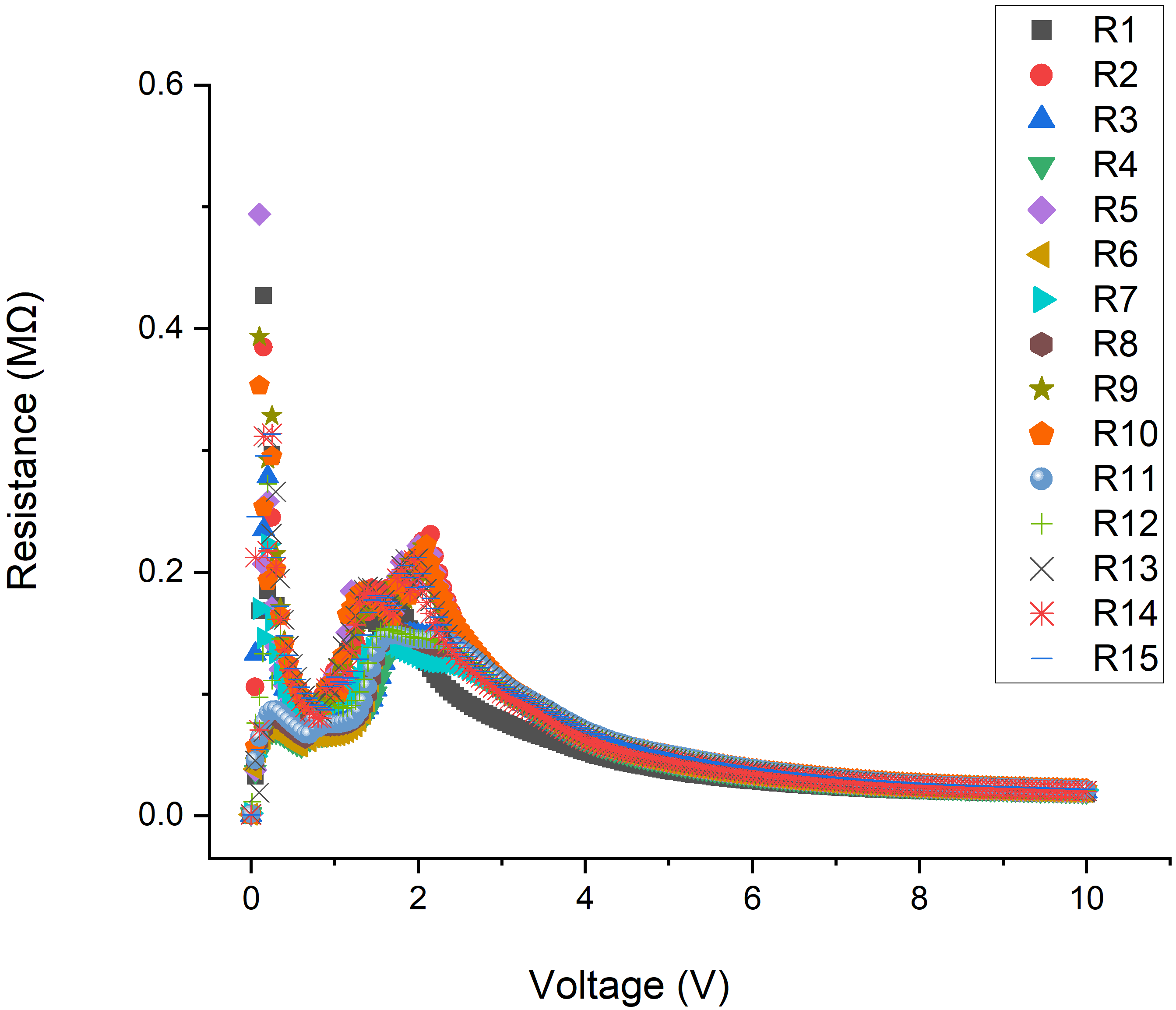}\label{fig:Ra10}}
\subfigure[]{\includegraphics[width=0.49\textwidth]{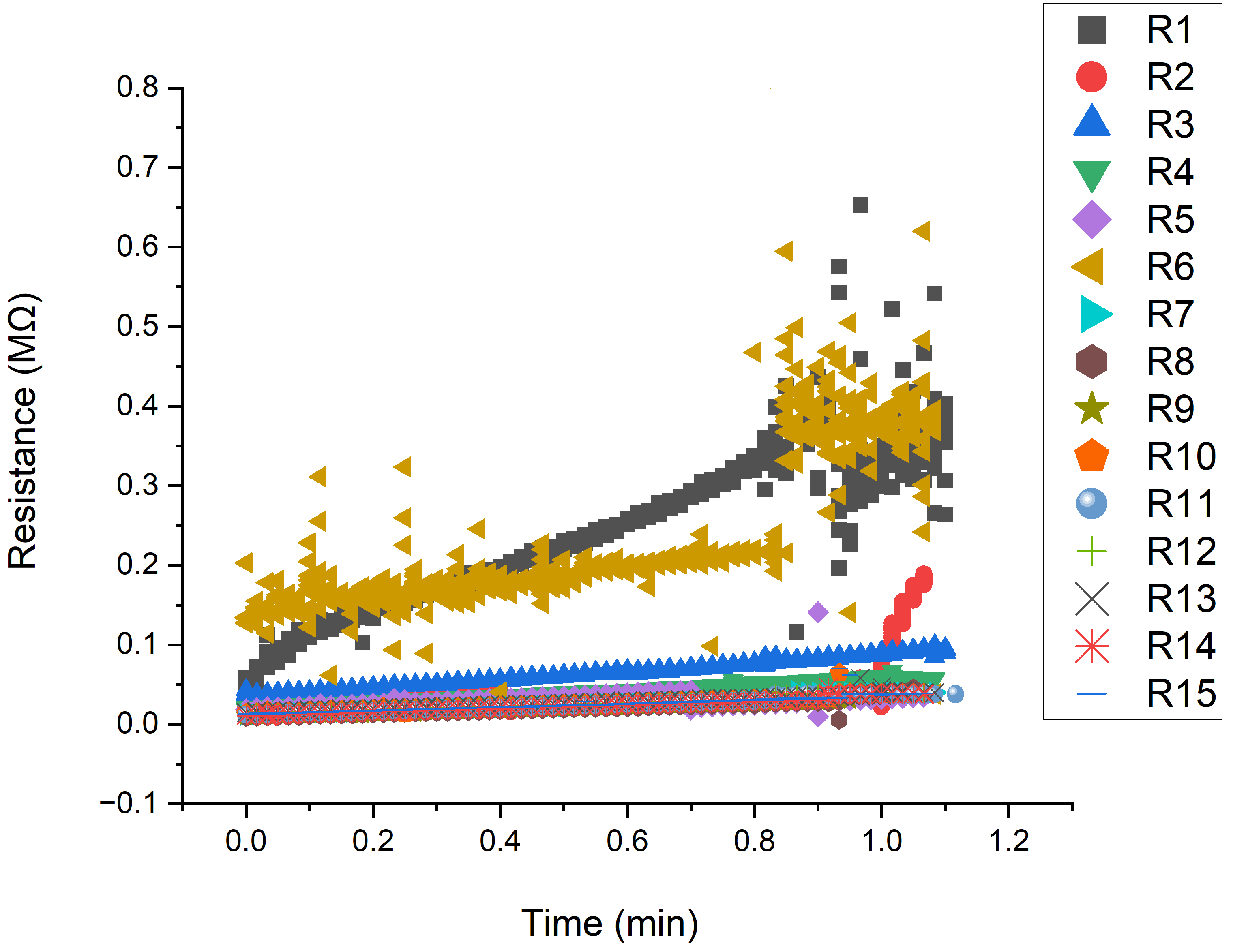}\label{fig:Rb10}}
\caption {(a)~Repeated resistance diagrams of sample A, during applying DC stimulation from \SIrange{0}{10} {\volt} to sample A, each step 50~mV.
(b)~Repeated resistance diagrams of sample B, during applying DC stimulation from \SIrange{0}{10} {\volt} to sample A, each step 50~mV}
\end{figure}

Second, a pre-synaptic spike (10~V stimulation) was applied just to sample “A” and the resistance of both samples was measured (mimicking the ringing of the “bell”), shown in Figs.~\ref{fig:Ra10} and \ref{fig:Rb10}.

\begin{figure}[!tbp]
\subfigure[]{\includegraphics[width=0.49\textwidth]{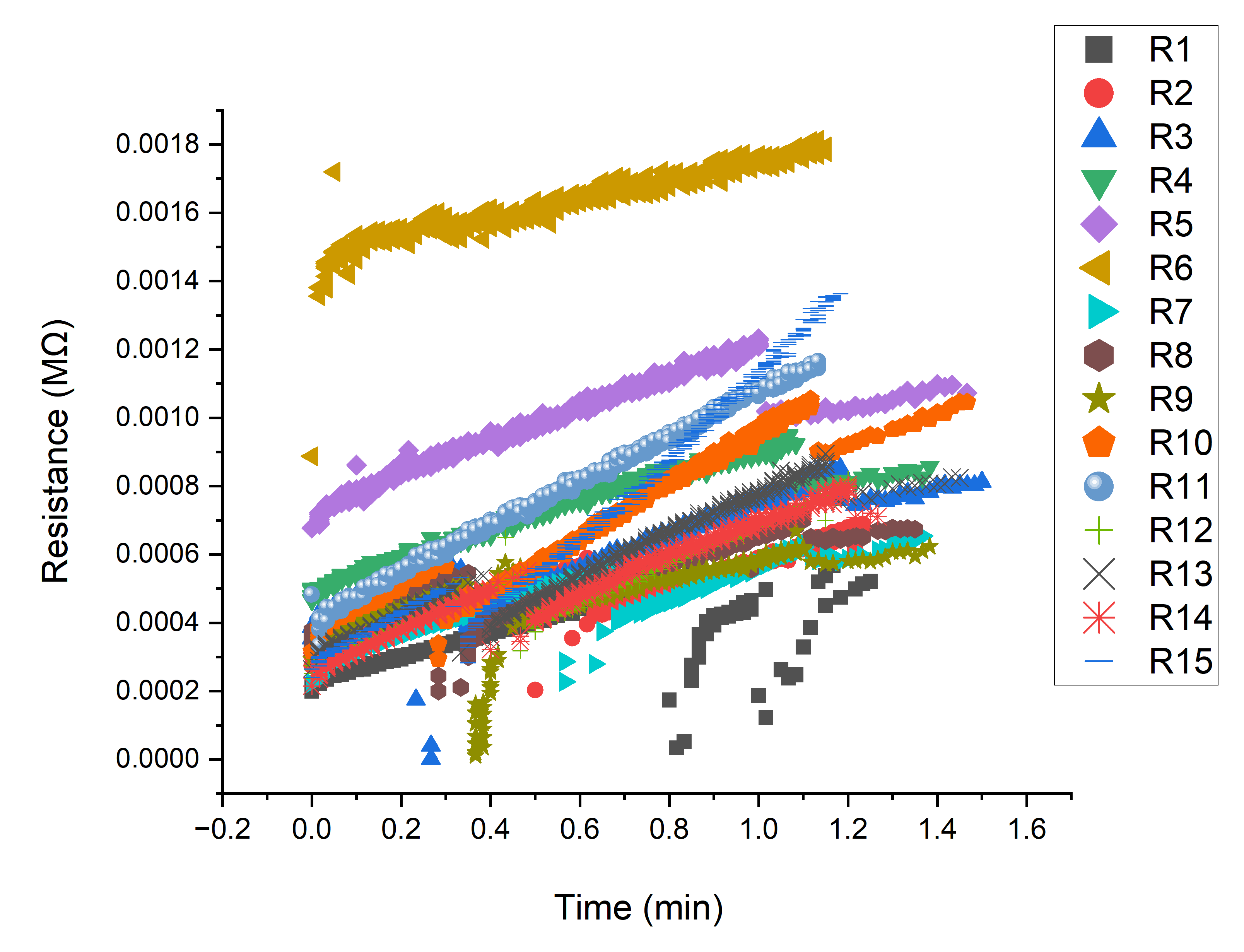}\label{fig:Ra3}}
\subfigure[]{\includegraphics[width=0.49\textwidth]{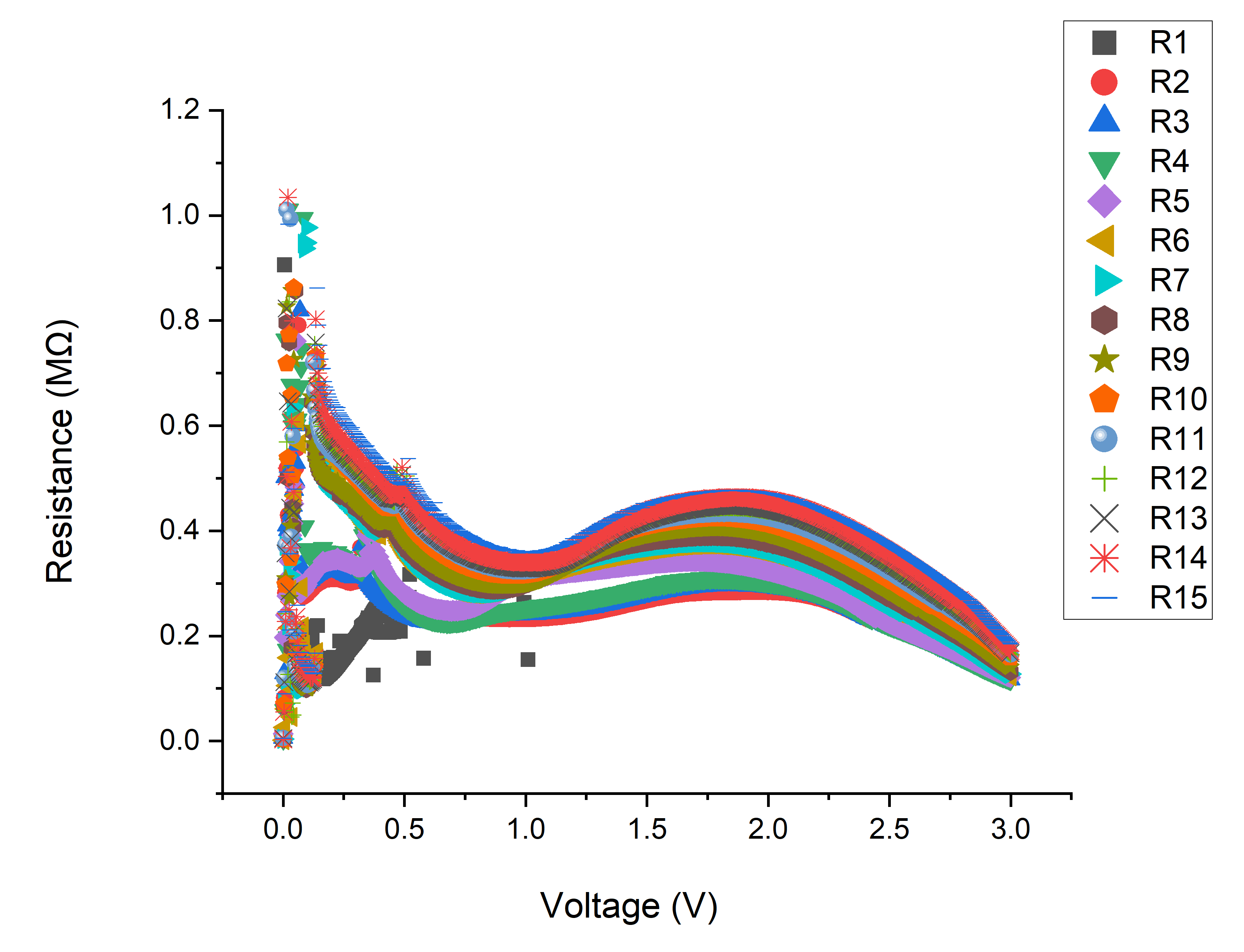}\label{fig:Rb3}}
\caption {(a)~Repeated resistance diagrams of sample A, during applying DC stimulation from \SIrange{0}{3} {\volt} to sample B, each step 10~mV.
(b)~Repeated resistance diagrams of sample B, during applying DC stimulation from \SIrange{0}{3} {\volt} to sample B, each step 10~mV.
}
\end{figure}

\begin{figure}[!tbp]
\includegraphics[width=\textwidth]{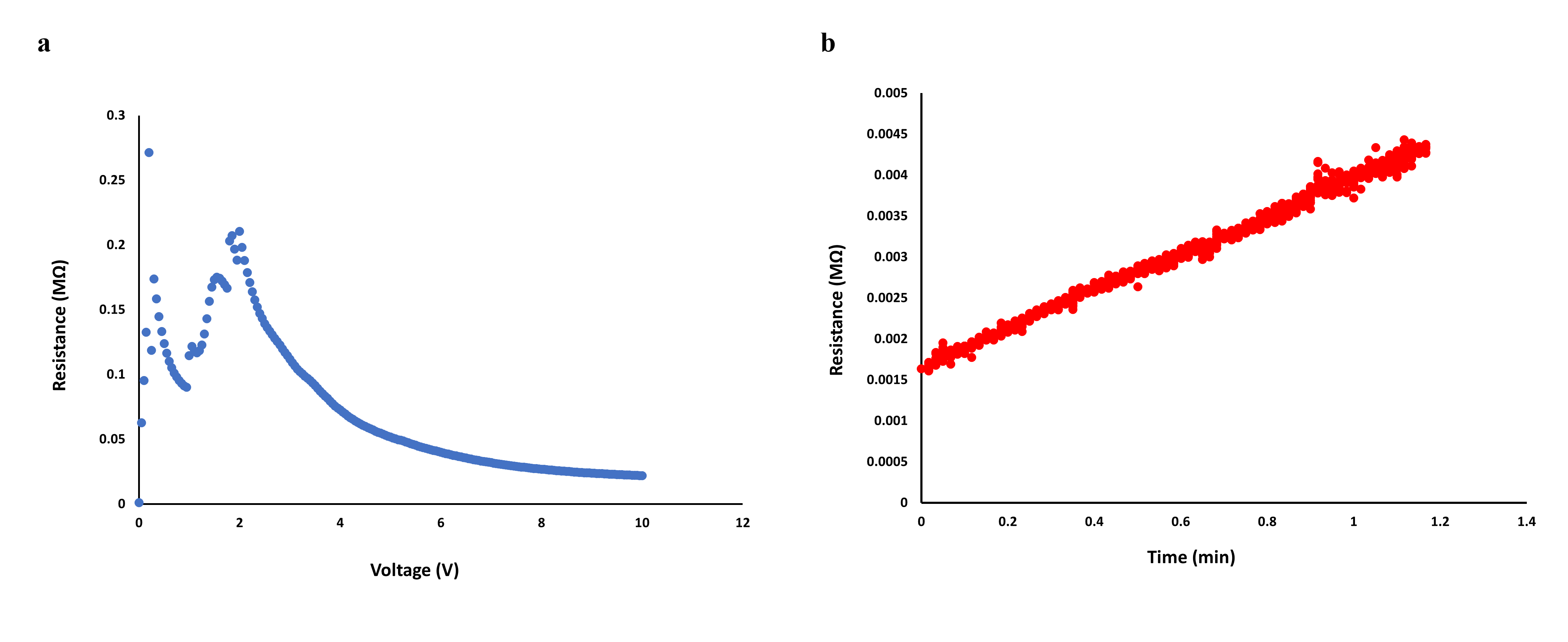}
\caption {Resistance diagram of sample A (left) and B (right), during applying DC stimulation from \SIrange{0}{10} {\volt} to sample A, each step 10~mV (just bell)}
\label{fig:bell}
\end{figure}

Third, a post-synaptic spike (3~V stimulation) was applied just to sample "B" and the resistance of both samples was measured (It acts as Food), Figs~\ref{fig:Ra3} and \ref{fig:Rb3}.
Salivation is equated to the dropping of the resistance of sample “B”. This cycle was repeated 15 times. Then with just applying 10~V stimulation to sample “A” like a bell, as we expected from Pavlovian conditioning, the resistance of sample “B” dropped down (Fig.~\ref{fig:bell}).

\begin{table}[!tbp]
\caption{Resistance of sample ``B” before performing Pavlovian conditions is about 1.6~M$\Omega$. The resistance of ``B'' during each step of the stimulation is shown in the table. The resistance of ``B'' after 15 cycles of stimulation is   0.038~M$\Omega$.}
\label{tab01}
\scriptsize
\begin{tabular}{ | m{2cm} | p{0.3cm}| m{0.31cm} | m{0.31cm} | m{0.31cm} | m{0.31cm} | m{0.31cm} | m{0.31cm} | m{0.31cm} | m{0.31cm} | m{0.31cm} | m{0.31cm} | m{0.31cm} | m{0.33cm} | m{0.36cm} | m{0.36cm} |} 
 \hline
Cycles & 1 & 2 & 3 & 4 & 5 & 6 & 7 & 8 & 9 & 10 & 11 & 12 & 13 & 14 & 15 \\ \hline
R$_B$ (M$\Omega$) after stimulating sample A till 10~V (Bell) & 0.26 & 0.18 & 0.09 & 0.06 & 0.03 & 0.04 & 0.04 & 0.04 & 0.04 & 0.04 & 0.04 & 0.04 & 0.04 & 0.038 & 0.038 \\ \hline
R$_B$ (M$\Omega$) after stimulating sample B till 3~V (Food) & 0.11 & 0.12 & 0.12 & 0.12 & 0.12 & 0.12 & 0.13 & 0.13 & 0.14 & 0.15 & 0.16 & 0.16 & 0.16 & 0.17 & 0.18 \\ \hline
\end{tabular}
\label{table}
\end{table}

Table~\ref{table} shows the resistance of sample “B” before performing Pavlovian conditioning, during each step and after finishing cycles.
As it can be seen, the resistance of sample “B” dropped down after 15 cycles of experiments from 1.6~M$\Omega$ to 38~K$\Omega$.
Based on this achievement, it can be concluded that Pavlovian learning can be implemented in colloids.

\begin{figure}[!tbp]
\includegraphics[width=\textwidth]{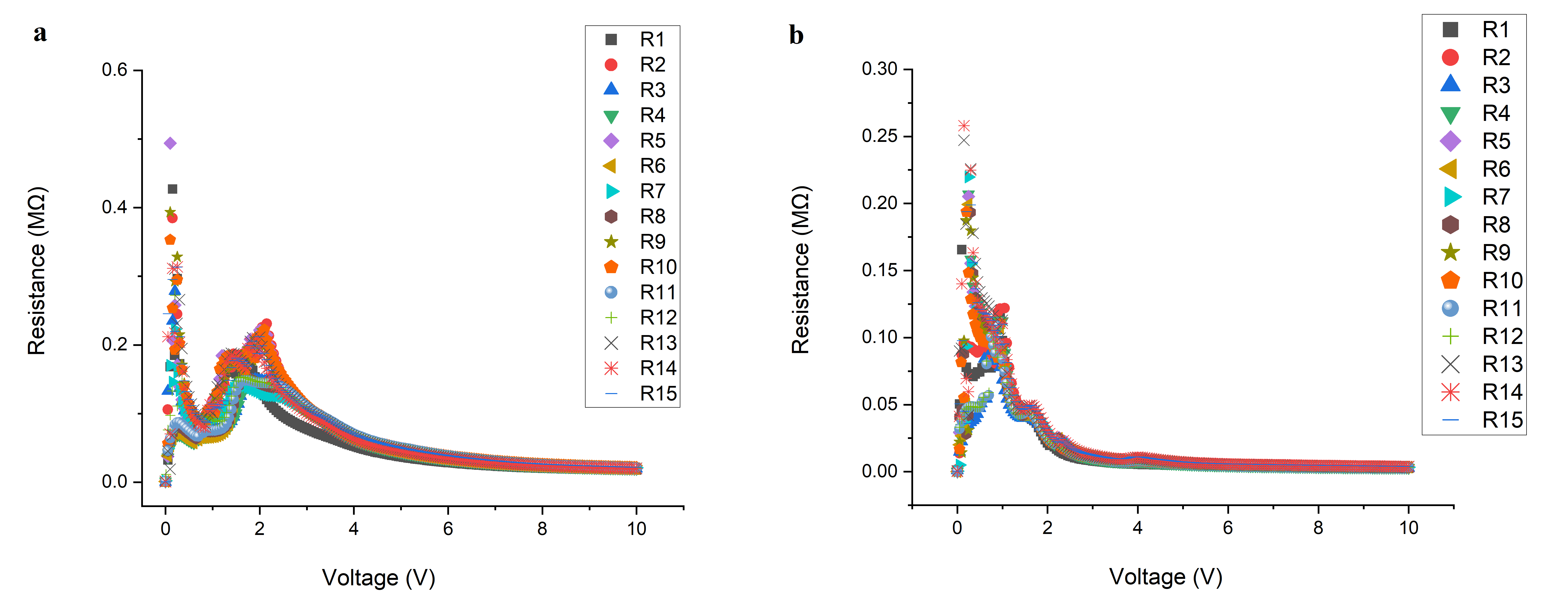}
\caption {(a) Repeated measurement resistance of colloid with DC stimulation from \SIrange{0}{10} {\volt} to sample A in the ambient atmosphere. (b) Repeated measurement resistance of colloid with DC stimulation from \SIrange{0}{10} {\volt} to sample A in the Nitrogen atmosphere and at $12^\circ C$}
\label{fig:comparison}
\end{figure}

In order to evaluate the effects of Brownian movements and oxygen interactions, we repeated applying the 10~V DC stimulation to sample "A" (mimicking the ringing of the “bell”) under Nitrogen gas and at $12^\circ C$. A total of 15 tests were conducted. Figure~\ref{fig:comparison} shows the results of this experiment and compares them with the previous tests in an ambient atmosphere. There is no significant difference between these two atmospheric conditions. Consequently, Brownian movements and oxygen interactions do not appear to affect the learning process of the synthesized colloid.

\section{Discussion}
\label{discussion}

In experimental laboratory conditions we demonstrated that it is possible to implement Pavlovian conditional reflexes with two geometrically constrained colloid volumes. Further cascading of the Pavlovian circuits and their integration into a large scale colloid neuromorphic circuits will require flexible containers and connections. Amongst a plethora of possible scenarios we can select two most feasible approaches. First, we can constrain the colloids into liquid marbles~\cite{aussillous2001liquid}. Liquid marbles (LMs) are droplets of liquid coated with hydrophobic power. The LMs behave as soft bodies and are easy to manipulate, they can be arranged into arbitrary geometries. Moreover, LMs can be made with a conductive coating, e.g. Ni particles, thus allowing to make connections between the colloid cargo by simply placing the LMs in physical touch with each other. Feasibility studies on LMs with Belousov-Zhabotinsy cargo, see e.g.~\cite{adamatzky2022sensing}, shown that computing circuits made of LMs are robust and can function for weeks. Second approach could be in producing the physical structure of a neuromophic device as a network of channels filled with colloid inside a polydimethylsiloxane volume, see e.g.~\cite{kheirabadi2022learning,chiolerio2020tactile}. Such enclosure will allow for flexibility, robustness and tolerance to environmental conditions.

\section{Conclusion}
\label{conclusion}

As the main unit of the nervous system, synapses transmit information between neurons, as well as process learning, memory, and forgetting.
In this paper, we demonstrate the synaptic properties of ZnO-based colloids. The colloid showed a comprehensive synaptic function in the Pavlovian associative learning process in the human brain.
Artificial intelligence computation, including learning-driven functions, can be achieved using this approach to neuromorphic computing.

\section{Acknowledgements}
This project has received funding from the European Union’s Horizon 2020 research and innovation programme FET OPEN ``Challenging current thinking” under grant agreement No 964388.


\end{document}